\documentclass{Interspeech2024}

\usepackage{graphicx, amsfonts, ,amsthm,amssymb,mathtools,float,hyperref,url, xcolor, makecell, booktabs, caption, subcaption, multirow,amsmath,enumitem}

\interspeechcameraready

\title{Speech After Gender: A Trans-Feminine Perspective on Next Steps for Speech Science and Technology}

\name[affiliation={1}]{Robin}{Netzorg}
\name[affiliation={2}]{Alyssa}{Cote}
\name[affiliation={3}]{Sumi}{Koshin}
\name[affiliation={4}]{Klo Vivienne}{Garoute}
\name[affiliation={1}]{Gopala Krishna}{Anumanchipalli}

\address{
  $^1$University of California, Berkeley, USA\\
  $^2$Alyssa's Voice Training, USA
  $^3$SumianVoice, Australia
  $^4$TransVoiceLessons, USA}
\email{robert\_netzorg@berkeley.edu, gopala@berkeley.edu}


\begin{document}

\maketitle


\begin{abstract}
    
    As experts in voice modification, trans-feminine gender-affirming voice teachers have unique perspectives on voice that confound current understandings of speaker identity. To demonstrate this, we present the Versatile Voice Dataset (VVD), a collection of three speakers modifying their voices along gendered axes. The VVD illustrates that current approaches in speaker modeling, based on categorical notions of gender and a static understanding of vocal texture, fail to account for the flexibility of the vocal tract. Utilizing publicly-available speaker embeddings, we demonstrate that gender classification systems are highly sensitive to voice modification, and speaker verification systems fail to identify voices as coming from the same speaker as voice modification becomes more drastic. As one path towards moving beyond categorical and static notions of speaker identity, we propose modeling individual qualities of vocal texture such as pitch, resonance, and weight.
\end{abstract}

\section{Introduction}

Speech systems are brittle. Historically, unseen speakers, accent \& dialect variation, and different microphone conditions can greatly affect the performance of speech systems across a variety of different tasks \cite{Trinh2022Elastic, Feng_2023}. In recent years, with the advent of large models, this brittleness has been reduced. Large foundation models can now robustly capture differences between speakers, environments, and recording devices \cite{ju2024naturalspeech, vyas2023audiobox}. As accuracy across diverse tasks improves, the question naturally arises of whether these current models are sufficiently robust to the variability of the human voice.

There are few in the world more aware of the variability of voice than trans and gender-diverse individuals. In their daily lives, trans and gender-diverse individuals regularly engage with the fact that listeners come to conclusions about the gender of an individual's voice in less than a second \cite{lavan2023time}, often resulting in gender judgements that override other aspects of presentation. Since feminizing hormone replacement therapy does not affect voice, trans women, for example, seeking to feminize their voices must either undergo surgical procedures or behavioral modification. As such, whether out of personal interest or necessity, trans women experiment daily with the production and perception of voice, becoming all too familiar with the nuances of vocal identity.

Despite its importance, gender-affirming voice modification is a notoriously difficult task within the trans and gender-diverse community \cite{bushgavtsurvey2022}, with experienced clinicians being rare \cite{brennan2022genderdifferences} and gender-affirming voice training only just starting to be studied by the scientific community \cite{adler2018voice}. Due to these limitations, many trans individuals seek training from informal online sources, such as tutorial videos or private coaches \cite{bushgavtsurvey2022}, which we will group together as the Informal Trans Voice Training Community (IVTC). Spanning the Internet, the IVTC is a broad community across Discord servers, YouTube channels, and Reddit forums. These sources primarily consist of private individuals, many of whom are trans themselves, teaching techniques they developed or adapted from other sources for their own voice training, often without institutional or formal training \cite{huff2022modern}. Each subcommunity uses similar language to describe perceptual qualities of voice, such as resonance and weight, but lack of formalized research leads to disagreement upon the precise importance and definitions of these qualities.

Exploring perspectives from the Informal Trans Voice Community, we bring together three gender-affirming voice teachers from different subcommunities of the IVTC to record and analyze audio clips demonstrating the versatility of voice, creating the Versatile Voice Dataset (VVD). With the VVD, we further demonstrate how current approaches to speaker modeling fall short when drawing inferences from diverse voices produced by a single speaker. 

With training, the human voice is incredibly flexible, and the limits of intra-speaker vocal flexibility are unknown. The diversity of single speaker voices we present in this work pose a thrilling challenge and new direction for the speech community. By centering the experiences of the transgender and gender-diverse community, we have the opportunity to model and understand the full capabilities of the human voice.

Our contributions are the following:
\begin{enumerate}
    \item The creation of the Versatile Voice Dataset, a collection of 3 trans-feminine voice teachers reading the 6 CAPE-V sentences in 27 different configurations of pitch, resonance, and weight.

    \item A perceptual study of the VVD, where we explore how well non-expert and expert listeners can perform speaker verification on novel and diverse voices.

    \item An evaluation of current speaker embeddings on gender and speaker identity modeling, and possible next steps via modeling vocal texture.
\end{enumerate}

\section{Related Work}

\subsection{Speaker Identity Modeling}

Modeling speaker identity is a fundamental task in speech processing, with applications to downstream tasks like voice conversion \cite{lian2022robust}, speaker recognition \cite{xvector}, diarization \cite{dawalatabad2021ecapadiar}, and anonymization \cite{panariello2024speaker, fang2019speaker}. While there is much work to produce new voices and manipulate speaker or style embeddings in voice synthesis and modification \cite{ju2024naturalspeech, guo2022prompttts}, voices are often treated as unique identifiers of speakers, with intra-speaker modification contained to emotive speech \cite{ju2024naturalspeech}. Descriptions of speaker identity are often broad, demographic categories such as 'Male/Female', 'Child/Adult/Elderly', or 'Healthy/Pathological'. There has been some recent work attempting to produce speaker embeddings that describe voices along perceptual axes \cite{netzorg2023interpretable}, but this prior work is based on cisgender voices, which omits a large swath of perceptual diversity.


\subsection{Trans and Gender-Expansive Speech}

Understanding the similarities and differences between trans, gender-expansive and cis-normative speech has gained interest in recent years. Prior work has noted that trans and gender-expansive individuals tend to use more diverse vocabulary and varied scalar ratings when describing the gender of an individual's voice \cite{brennan2022genderdifferences, hope2022gender}. These differences in perception are primarily attributed to trans and gender-expansive individuals' use of voice to present their gender identities.

Datasets that focus on exploring the acoustic properties of trans and gender-expansive speech are only just starting to be developed. A Palette of Voices for Transmasculine Individuals \cite{dolquist2023palette} and the Mid-Atlantic Gender-Expansive Speech Corpus \cite{hopenonbinary2023} are, to the best of our knowledge, the only two publicly available datasets of trans and gender-expansive speech currently available. While more exist, they are usually regional and privately held to protect the identity of trans and gender-expansive individuals \cite{merritt2022perceptual}. Analyses of trans and gender-expansive datasets focus on exploring inter-speaker acoustic variability across acoustic measures that correlate with gender, such as pitch or formants. 

While prior datasets are designed to explore inter-speaker variability, we are primarily interested in exploring intra-speaker variability, as seen through the lens of gender-affirming voice modification, in order to overcome the limitations of existing speaker identity models.

\section{Versatile Voice Dataset}

\begin{table*}[t]
    \centering
    \begin{tabular}{c|c|c|c|c}
         Speaker Id & Pronouns & Accent & Voices Produced & Microphone \\
         \hline
         001 & She/Her & Californian American & 27 & AT2020  \\
         002 & Any & Australian & 25 & RZ19-03450100 \\
         003 & She/Her & Californian American & 27 & AKG P120 \\
    \end{tabular}
    \caption{Speaker Information for the Three Speakers in the Versatile Voice Dataset.}
    \label{tab:spk_info}
\end{table*}

In this section, we present the Versatile Voice Dataset (VVD), the first public dataset of speakers drastically modifying the perceived speaker identity of their own voice. We first define the terms pitch, resonance, and weight as voice teachers within the ITVC understand them, and then describe the collection process for the VVD. Samples from the VVD are provided online\footnote{https://berkeley-speech-group.github.io/VersatileVoiceDataset/}. 

\subsection{Pitch, Resonance, and Weight}\label{sec:prw}

As these qualities have been found to be extremely important in gender perception \cite{markova2016age}, pitch, resonance and weight are the most commonly used concepts in the ITVC to teach behavioral voice modification. Connecting the concepts to acoustic and physical phenomenon, pitch is understood as the fundamental frequency, resonance as the acoustic qualities of the overall vocal tract shape (pharyngeal, oral, and nasal) \cite{kent1993vocal}, and weight as the sound quality associated with the vocal fold vibratory mass, which correlates with the closed quotient \cite{huff2022modern}.

While pitch, resonance, and weight as concepts are mostly agreed upon across the ITVC, there is more ambiguity in these terms when it comes to precise definitions and perceptual descriptions, especially for weight. Speakers 001 and 003 describe weight as a ``buzzy'' sound, whereas 002 describes it as more of a ``rumble'', only sounding buzzy under high adduction, high resonance configurations. These disagreements serve as exciting stepping stones for future perceptual studies.

It is important to note that many teachers in the ITVC highlight different aspects of voice than those typically emphasized in speech language pathology. Rather than an emphasis on oral cavity resonance and pitch as in speech language pathology \cite{adler2018voice}, many in the ITVC choose to focus on pharyngeal resonance, and use the concept of weight to balance pitch modification \cite{huff2022modern}.

\subsection{Data Collection}\label{sec:data_col}

Along self-interpreted configurations of low, medium, and high for pitch, resonance, and weight, 3 trans-feminine gender-affirming voice teachers produced up to 27 voices. With each voice configuration, the voice teacher read the six CAPE-V sentences aloud \cite{capev2009}, for a total of 14.5 minutes of speech. Summary of speaker information is provided in Table \ref{tab:spk_info}.

As the ITVC is broad, the three voice teachers had not interacted before the collection of this data and had not heard each others' voices. The ambiguous instructions of low, medium, and high configurations led to each teacher interpreting the terms and task slightly differently. Speaker 001, for example, speaks with a yawn-like voice in the low-low-low configuration, while Speaker 002 and 003 speak in more natural-sounding voices. Similarly, different interpretations of weight lead to voice variation, reflected in Speaker 002 omitting med-high-high and high-high-high configurations to avoid vocal strain. Without an external and consistent measurement of pitch, resonance, and weight, these discrepancies lead to difficulties in drawing conclusions on inter-speaker comparisons such as naturalness of speech across configurations. 

Standardizing the sampling rate to 16kHz is the only pre-processing performed on the VVD clips. No other pre-processing was performed in order to preserve each speaker's interpretation of pitch, resonance, and weight. As all audio clips were provided by authors of this paper, no ethics approval was necessary.

\section{Limitations of Current Speaker Embeddings}

Here, we explore how intra-speaker variation in the VVD affects state-of-the-art ECAPA-TDNN speaker embeddings \cite{ecapatdnn2020} from SpeechBrain \cite{speechbrain} and NeMo \cite{nemo2024} on two tasks in speaker modeling: Gender Classification and Speaker Verification.

\subsection{Gender Classification}

In many speech models, gender information is used as a proxy for vocal texture, converting the high-dimensional space of vocal acoustics into a small number of categories. The utility of these categories is based on prior work on gendered differences in speech \cite{huff2022modern} and the ability of gender classification models to accurately infer gender categories from speech \cite{ellis2023nonbinaryclass}. Using Scikit-Learn's default parameter settings, Random Forest models predicting speaker gender in VCTK \cite{yamagishi2019vctk} from either NeMo or SpeechBrain ECAPA-TDNN embeddings achieve high accuracies of 99.1\% and 97.0\% accuracy on the test split of VCTK, respectively.

\begin{figure}[t]
    \centering
    \includegraphics[scale=0.5]{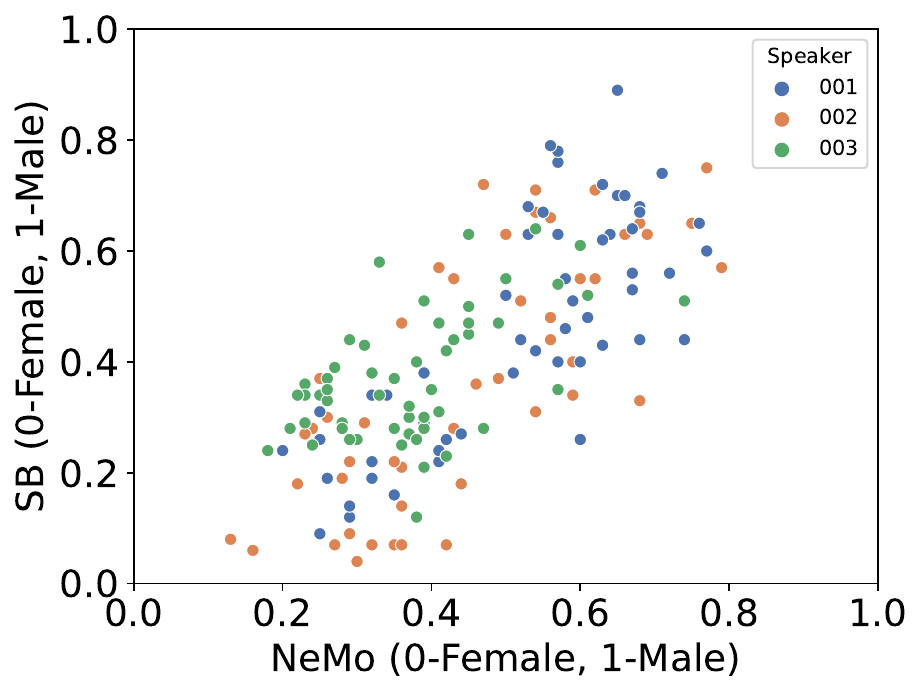}
    \caption{Predicted gender identity by NeMo's ECAPA-TDNN model compared with predicted gender identity by SpeechBrain's ECAPA-TDNN model across all speakers.}
    \label{fig:gender_class}
\end{figure}

Using gender as a proxy, however, begins to lose its predictive power with more vocal diversity. Visualizing the gender classifiers' predictions over the VVD in Figure \ref{fig:gender_class}, we see that the predictions across voices span almost the entire range of 0 (Female) to 1 (Male) for both models and all speakers.

\begin{table}[t]
\centering
\begin{tabular}{c|c|c|c}

Pitch & Resonance & NeMo & SpeechBrain \\
\hline
\multirow{3}{*}{low} & low & 0.61 & 0.61 \\  
                     & med & 0.60 & 0.46 \\  
                     & high & 0.51 & 0.38 \\ \hline
\multirow{3}{*}{med} & low & 0.58 & 0.56 \\  
                     & med & 0.50 & 0.41 \\  
                     & high & 0.40 & 0.29 \\ \hline
\multirow{3}{*}{high} & low & 0.52 & 0.55 \\ 
                      & med & 0.48 & 0.40 \\ 
                      & high & 0.32 & 0.25 \\ 
\end{tabular}
\caption{Probability of a given voice being Male averaged across configurations of Pitch and Resonance. Std. Error $\leq0.02$ for all entries, as estimated via bootstrap.}
\label{tab:gender_manip}
\end{table}

Furthermore, differences in predicted gender are not arbitrary, but the result of the deliberate modifications made by the three speakers. Shown in Table \ref{tab:gender_manip}, pitch and resonance levels directly influence the gender prediction of the model, with both models predicting more feminine voices as both pitch and resonance increase. While current speaker embeddings do implicitly encode high-level perceptual qualities such as pitch and resonance, multiple configurations produce similar predictions, revealing an ambiguity between vocal configuration and binary gender.   

These exercises illustrate how, by summarizing vocal diversity with a small number of identity-based categories, gender obscures the richer acoustic space that affects voice perception. While helpful for modeling many voices, categorical notions of gender can result in unreliable and possibly harmful classifications to those who deviate from prototypical voices. Knowing that a voice is a woman's, for example, does not accurately identify the unique texture of that voice. Moreover, systems based on categorical gender can be manipulated by deliberate behavioral changes to voice, which we explore further in Section \ref{sec:sv}.

\subsection{Speaker Verification}\label{sec:sv}

\subsubsection{Equal Error Rate}

The goal of speaker verification is to determine whether or not two voice clips are created by the same speaker. Many of these methods are trained on cross-entropy loss alongside that encourage inter-class diversity and intra-class compactness, such as additive angular margin loss (AAM-Softmax) \cite{xiang2019margin}. Coupled with architecture design, state-of-the-art models like the ECAPA-TDNN models from SpeechBrain and NeMo are able to achieve low Equal Error Rates (EERs) on test sets, achieving EERs on VoxCeleb1 \cite{voxceleb2017} of 0.80\% and 0.92\%, respectively, being able to identify clips as originating from different speakers with remarkable accuracy.

This modeling approach to speaker verification fundamentally assumes that a speaker's vocal texture is static, which is violated by the speakers in the VVD. Illustrated in Table \ref{tab:eer}, the EER of speaker verification on VVD is remarkably higher than typical test performance, with EER on VVD being 21.52\% for NeMo and 29.00\% for SpeechBrain. This is not merely due to a distribution shift, as analysis on the test set of VCTK \cite{yamagishi2019vctk}, which was not used in either of the models' training, achieves similar a EER to the EER achieved on VoxCeleb1. 

\begin{table}[b]
    \centering
    \begin{tabular}{c|c|c}
        \textbf{Dataset} & NeMo & SpeechBrain  \\
        \hline
        VoxCeleb1 & 0.80 & 0.92 \\
        VCTK & 0.64 & 1.26 \\
        VVD & 21.52 & 29.00
    \end{tabular}
    \caption{Equal Error Rate (\%) across speaker verification models and datasets. VoxCeleb1 is in-distribution while VCTK and VVD are both out-of-distribution.}
    \label{tab:eer}
\end{table}

\begin{figure*}[t]
    \centering
     \begin{subfigure}[b]{0.45\textwidth}
         \centering
            \includegraphics[scale=0.35]{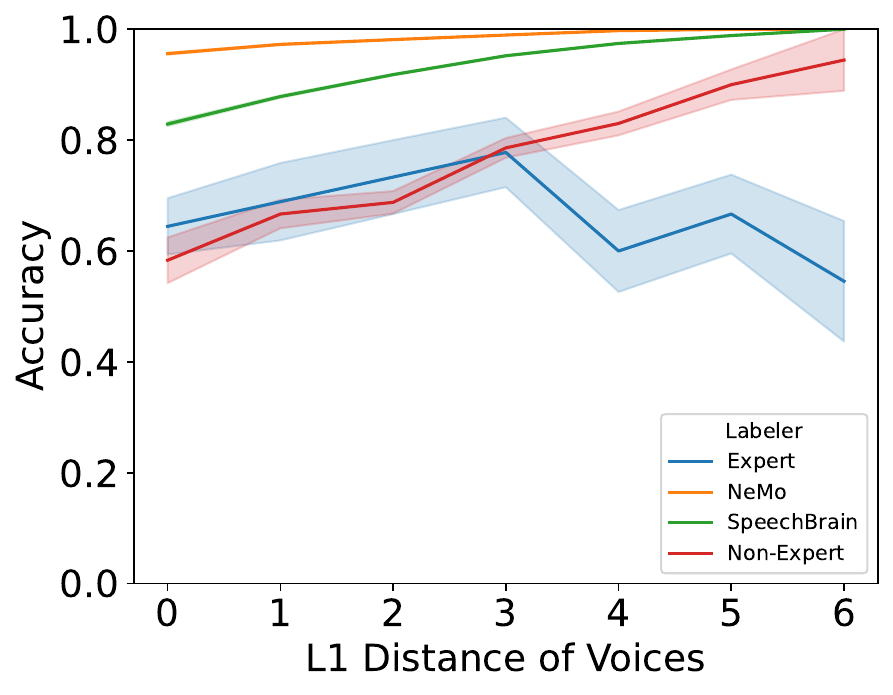}
         \caption{Different Speaker}
         \label{fig:diff_dist}
     \end{subfigure}
     \hfill
     \begin{subfigure}[b]{0.45\textwidth}
         \centering
            \includegraphics[scale=0.35]{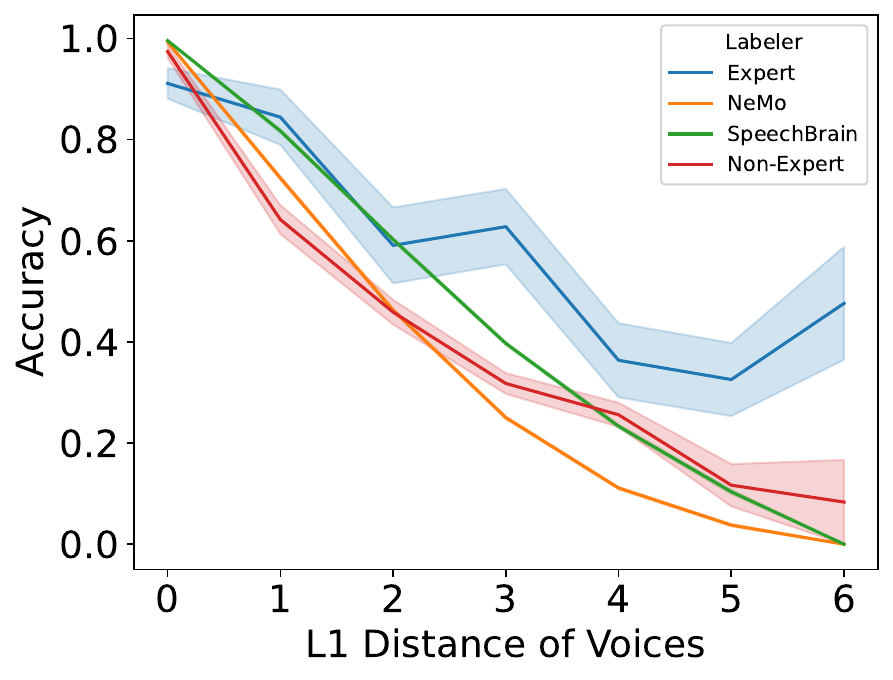}
         \caption{Same Speaker}
         \label{fig:same_dist}
     \end{subfigure}
    \caption{Accuracy ($\pm 1$ Std. Error) of Speaker Verification across L1 Voice Distance over voice clips produced by pairs of speakers.}
    \label{fig:perf_dist}
\end{figure*}

\subsubsection{Error Analysis and Difficulty}

The high-level EER results make it clear that speaker verification systems perform worse on the VVD; however, these results do not indicate how the systems misclassify voices, or how they compare to human listeners. We explore these questions here.

A unique quality of the VVD is the ability to directly measure the distance between voices, and then compare the performance of both automated speaker verification systems and human identification across said distance. Converting the levels of low, medium, and high to $0,1,2$, and then taking the L1 distance between two audio clips' three dimensional vectors, we map all pairs of voices onto a discrete space distance ranging between $0$ and $6$. For example, the distance between low-low-low and high-high-high would be a distance of $6$.

We perform two perceptual experiments with non-expert and expert listeners. As noted in Section \ref{sec:data_col}, the three voice teachers were unfamiliar with each others' voices. Therefore, for expert classifications, we asked each voice teacher (e.g. 002) to classify whether or not given pairs of clips from the other teachers (e.g. 001, 003) belonged to the same speaker. For non-expert listeners, we asked six workers with Master's Qualification on Amazon Mechanical Turk to perform the same task. Both non-experts and experts were given the same 200 tasks and instructions, whereby the clips in each task contained voices from a maximum of two speakers, and were balanced such that random guessing would result in a performance of 50\% accuracy. To measure classification accuracy for the SpeechBrain and NeMo ECAPA-TDNN models, we use both libraries' default threshold for classification evaluated on all possible pairs.

Averaging classification across all pairs and speakers, the achieved accuracies are: 59\% for non-experts, 63\% for experts, 78.47\% for SpeechBrain, and 77.89\% for NeMo. For human labelers, even those that are aware of the flexibility the vocal tract, speaker verification across diverse voices is a challenging task. Reporting the results split across pairs and distance in Figure \ref{fig:perf_dist}, a variety of interesting trends emerge. Across pairs coming from both different and same speakers, non-expert and model accuracies correlate. For pairs coming from different speakers, we see that accuracy increases as voices become distinct; for pairs coming from the same speaker, we see the opposite trend where accuracies trend towards $0$ as L1-Distance increases. 

While low same-speaker performance is expected as voices become distinct, due to the fundamental ignorance of speech modification by the considered models, the expert vs. non-expert performance comparison sheds light on the perception of voice. We see that, across both types of pairs, even experts struggle to tell if two voice clips are produced by the same speaker, performing at random when voices are most distinct. Despite non-experts being informed that a total of two speakers produced all voice clips in this task, non-expert performance diverges with that of experts, consistently rating voices with an L1-Distance greater than $3$ as belonging to different speakers. Additional prompting might increase non-expert same-speaker performance, but non-experts' assumptions of voice flexibility are similar to that of models: the flawed assumption that vocal texture is fixed.

\section{Next Steps: Modeling Vocal Texture}

Just as gender-diversity cannot be limited to a single label of ``cisgender'', ``non-binary'', or ``transgender'' (as notions of gender differ across cultures and time \cite{gill2024short}), categories of ``man'', ``woman'', and ``non-binary'' do not capture the full complexity of voice. Binary or categorical gender information, while helpful for defining a broad distribution over possible voices, results in large amounts of ambiguity concerning the specific texture that makes a voice unique. Moving away from a categorical notion of gender, models that center high-level perceptual qualities of voice, such as pitch, resonance, and weight, offer an opportunity to explicitly map the perceptual space of vocal texture. 

Providing intra-speaker labels of perceptual qualities, the VVD makes it possible to measure the performance of vocal texture models, such as the PQ-Representation \cite{netzorg2023interpretable}, which explicitly model resonance and weight. Evaluating the PQ-Representation's ability to rank intra-speaker variation of resonance and weight, we see in Table \ref{tab:percept_rank} that the PQ-Representation is unable to accurately rank the resonance of clips, performing slightly below random, and is only able to rank weight marginally better than ranking via Praat's harmonics-to-noise ratio (HNR) estimation. While prior work shows that the PQ-Representation accurately fits typical voices, the diversity of voices in the VVD leads to worse performance.  

To encourage further modeling of vocal texture, we propose the three dimensional vector of averages of F0, averages of the first through third formants, and averages of the HNR as a baseline measurement of the VVD's vocal qualities, achieving initial ranking accuracies of 91.7\%, 71.5\%, and 62.3\% for pitch, resonance, and weight respectively.

\begin{table}[b]
    \centering
    \begin{tabular}{c|c|c}
         - & Resonance & Weight \\
         \hline
         PQ-Representation &  49.5\% & \textbf{62.5\%} \\
         Avg. Formant-HNR & \textbf{71.5\%} & 62.3\% 
    \end{tabular}
    
    \caption{Accuracy of correctly ranking pairs of audio clips with different resonance or weight levels, as measured by the PQ-Representation and avg. formant and HNR. Mean Pitch ranking as provided by Praat achieves 91.7\% accuracy.}
    \label{tab:percept_rank}
\end{table}

\section{Conclusion}

The Versatile Voice Dataset demonstrates that there is a section of the population for whom static and categorical models of speaker identity will fail, but the shortcomings of these models provide a novel opportunity to advance the study of speech. Currently, we are unaware of the limits of behavioral voice modification, what a model of vocal tract flexibility would entail, or how to learn modification-robust speaker embeddings.  We are excited for the countless research questions behavioral voice modification poses to the speech community, and the future insights that may emerge through collaborations with those often outside of the traditional academic milieu.   

\section{Acknowledgements}

This work was supported by the UC Noyce Initiative, Society of Hellman Fellows, NSF, NIH/NIDCD and the Schwab Innovation Fund.

\bibliographystyle{IEEEtran}
\bibliography{mybib}

\end{document}